\documentclass[nofootinbib,preprint,amsmath,amssymb,showpacs,showkeys,aps]{revtex4-1}
\usepackage{graphicx}
\usepackage[utf8,latin1]{inputenc}
\usepackage{dcolumn}
\usepackage{mathrsfs}
\usepackage[overload]{empheq}
\usepackage{cases}
\usepackage{setspace}
\usepackage{bm}
\usepackage{bigints}
\usepackage{hyperref}
\usepackage{amsmath}
\usepackage[T1]{fontenc}
\newcommand{\qm}[1]{``#1''}
\usepackage{hyperref}
\hypersetup{colorlinks, linkcolor={red},citecolor={blue},urlcolor={blue}}  

\newcommand{\bsubeqs}{\begin{subequations}}
\newcommand{\esubeqs}{\end{subequations}}

\newcommand{\dd}{{\rm d}}
\newcommand{\OO}{{\rm O}}
\newcommand{\OR}[1]{{\rm O}\left(r^{#1}\right)}
\newcommand{\RS}{R_{\rm S}}
\newcommand{\LP}{\ell_{\rm P}}
\def\nn{\nonumber}

\begin{document}

\title[Quantum Schwarzschild geometry in effective-field-theory models
of gravity]{Quantum Schwarzschild geometry \\ in effective-field-theory models
of gravity}

\author{Emmanuele Battista}\email{emmanuelebattista@gmail.com;emmanuele.battista@univie.ac.at}

\affiliation{Department of Physics, University of Vienna, Boltzmanngasse 5, A-1090 Vienna, Austria}

\date{\today}

\begin{abstract}
The Schwarzschild geometry is investigated  within the context of effective-field-theory models of gravity. Starting from its harmonic-coordinate expression, we derive the  metric in standard coordinates by keeping  the leading one-loop quantum contributions in their most general form. We examine the metric horizons and the nature of the  hypersurfaces having constant radius; furthermore,  a possible energy-extraction process which violates the null energy condition is described, and both timelike and null geodesics are  studied. Our analysis shows that there is no choice of the sign of the constant parameter embodying  the quantum correction to the metric which leaves all the features of the classical Schwarzschild solution almost unaffected.

\end{abstract}

\maketitle

\section{Introduction}
\label{sec:intro}

Scientific laws span several length or energy regimes, which vary from the cosmological  to the   particle physics domains. In each energy range,  physical models can be usually investigated  independently of the laws at all  energies, since the scales which are smaller or larger than those characterizing  the phenomenon under study can be often set  to zero or  infinity, respectively, to obtain the correct description of the process. This represents the basic principle of effective field theories (EFTs) \cite{Weinberg2009,Isidori2023}, whose  techniques are widely exploited in physics. A  well-known  example is furnished by the standard model of particle physics, which is supposed to represent  the low-energy counterpart of a more fundamental theory \cite{Willenbrock2014,Brivio2017}. In addition, Newtonian mechanics can be seen as the effective model of special relativity in the limit of small energies and low velocities.

General relativity (GR) naturally fits into the  EFT paradigm and forms a well-behaved quantum theory at energies far below the Planck mass \cite{Donoghue1994,Donoghue1995c,Burgess2003,Donoghue2012a,Donoghue2017a,Battista-book2017,Burgess2020,Donoghue2022}. In this setup, gravitational interactions  can be readily organized into an energy expansion and the low-energy quantum effects can be separated from those stemming from the (unknown) high-energy regime of the theory. Following the guidelines of EFT, GR is characterized by an effective action  whose various terms correspond to different energy scales. In fact,  apart from  the Einstein-Hilbert part and the minimally coupled matter contributions, it  includes all higher-derivative pieces which are compatible with general covariance. In this way, the traditional renormalization issues which plague GR \cite{tHooft1974,vanNieuwenhuizen1976,Goroff1985,DeWitt2003,Hamber2009,Battista-book2017}  can be avoided.  Indeed, all the ultraviolet divergences can be absorbed into the phenomenological coefficients appearing in the effective action provided that  a regularization scheme which does not break general covariance is adopted. 

In the EFT scenario, only a finite number of parameters of the action need to be considered at each order in the energy expansion. Our ignorance  of the full theory of quantum gravity is thus  hidden in a few quantities, which can then be used to make reliable predictions. The leading, i.e., one-loop, long-distance quantum corrections  arise from the Einstein-Hilbert sector of the effective action and represent the first quantum modifications affecting gravitational processes. These contributions are independent of the high-energy completion of quantum gravity, since they are related to the propagation of  massless particles, and produce nonlocal contributions in coordinate space (or,  equivalently,  nonanalytic terms  in momentum space)  to the vertex functions and propagators. 

EFT tools have been  employed to derive one-loop  quantum corrections to the Newtonian gravitational potential \cite{Donoghue1994PRL,Donoghue1994,Bjerrum-Bohr2002} (see also Refs. \cite{Muzinich1995,Akhundov1996,Khriplovich2002,dePaulaNetto2021,Frob2021}), and the Schwarzschild, Kerr, Reissner-Nordstr\"{o}m, and Kerr-Newman metrics  \cite{Donoghue2001,Bohr2003,Kirilin2006,Calmet2017,Calmet2019}. Subsequently, many applications of these results have been considered in the literature.  Indeed, EFT pattern has been used to  examine  the classical and quantum modifications to the location of  both stable and unstable Lagrangian points in the Earth-Moon-satellite and Sun-Earth-satellite three-body systems \cite{Battista2014a,Battista2014b,Battista2015a,Battista2015b,Battista-book2017,Tartaglia2018,Ansari2022}. Moreover, Refs. \cite{Bai2017,Battista2017a,Kim2022} deal with the quantum contributions to the bending of light around the Sun, Shapiro time delay, and  frame-dragging effect. The motion of three masses has been explored in Refs. \cite{Battista2020c,Alshaery2020,Singh2021,Abouelmagd2023}, while the dynamics involving purely classical post-Newtonian effects has been studied in Refs. \cite{Yamada2010b,Yamada2010,Ichita2010,Yamada2015,Asada2022}. Lastly, also topics like compact binary inspirals and the ensuing gravitational radiation  \cite{Jenkins2018,Liu2023}, and  the  Newtonian cosmology \cite{Mandal2022} have been analyzed  via the EFT paradigm. 

The calculations behind the abovementioned investigations have been performed by resorting to the traditional Feynman diagrammatic scheme and the underlying vertex rules of the gravitational Lagrangian. Recently,  a novel and more straightforward approach which permits to greatly simplify the evaluation of loop diagrams in GR settings  has been devised \cite{Bjerrum-Bohr2013}. It relies on the  spinor-helicity variables, on-shell unitarity methods, and Kawai-Lewellen-Tye (KLT) relations \cite{Bern1994,Bern2002,Bern2011}. By means of these modern tools, the authors of Ref. \cite{Bjerrum-Bohr2013} have proved that the spin-independent components of one-loop gravity amplitudes exhibit a universal character (i.e., their form does not depend on the type of interacting matter), and have also confirmed the analysis of the leading quantum corrections to  the Newtonian potential contained in  Ref. \cite{Bjerrum-Bohr2002}. Furthermore, one-loop modifications to the deflection angle of a beam of massless particles scattered off a massive scalar object (such as the Sun)   have also been worked out \cite{Bohr2015,Bohr2016,Chi2019} (note however that in this case there is a discrepancy between Refs. \cite{Bohr2015,Bohr2016} and \cite{Bai2017,Chi2019}).  The importance of  double-copy constructions and unitarity-based strategies is also witnessed by the research project conceived in Refs. \cite{Bjerrum-Bohr2018,Bjerrum-Bohr2021}, where it has been proposed a program to obtain classical GR results from scattering amplitude techniques. In particular, the aforementioned unitarity procedures can be exploited to perform a post-Newtonian and  post-Minkowskian analysis of GR, with important implications in the field of gravitational waves \cite{Vanhove2022} and the relativistic two-body problem \cite{Damgaard2021}.

In this paper, we study the main features of the quantum Schwarzschild geometry within the context of EFTs (for other approaches to the quantum Schwarzschild metric, see e.g. Refs. \cite{Bonanno2000,Bargueno2016,Ashtekar2018,Faraoni2020,Beltran-Palau2022,DAlise2023} and references therein). In our analysis, we keep one-loop contributions in their most general form and  explicitly construct the transformation of coordinates which permits to write the metric in Schwarzschild coordinates starting from its expression in harmonic coordinates, which are  usually employed in EFT calculations \cite{Bohr2003}. Moreover, we do not adopt from the very beginning the simplifying hypothesis $-g_{tt} = g^{rr}$, which on the contrary can be found e.g. in  Refs. \cite{Bonanno2000,Bargueno2016}. The plan of the paper is thus as follows. After having  derived the quantum  metric in Schwarzschild coordinates in Sec. \ref{Sec:Quantum-metric-standard-coord}, we examine  some applications in Sec. \ref{Sec:applications}, where we consider the horizons, null hypersurfaces, curvature invariants, and geodesic motion. Concluding remarks are finally given in Sec. \ref{Sec:Conclusions}.

\section{Quantum Schwarzschild metric in standard coordinates} \label{Sec:Quantum-metric-standard-coord}

In this section, we derive the expression of the Schwarzschild metric containing the leading long-range quantum corrections in standard coordinates $(ct,r,\theta,\phi)$. We begin our analysis with a brief review of the classical result  (see Sec. \ref{Sec:classical-case}). Then, in  Sec. \ref{Sec:Quantum-corrected-Schwarzschild-metric} we describe  the procedure to obtain the quantum metric starting from its harmonic-coordinate expression provided by the EFT scheme. 

\subsection{The classical Schwarzschild metric}
\label{Sec:classical-case}

The Schwarzschild metric is the unique static, spherically symmetric, asymptotically-flat  solution of the vacuum Einstein equations \cite{Wald,Weinberg1972}. It permits to describe the gravitational field outside a static spherical source or a  black hole.  In the so-called Schwarzschild (or standard or spherical) coordinates $x^\mu=(ct,r,\theta,\phi)$, it reads as
\begin{equation}
\dd s^2=g_{\mu \nu} \dd x^\mu  \dd x^\nu= - \mathcal{B}(r)\left(c^2{\rm d}t^2\right) + \mathcal{A}(r){\rm d}r^2 + r^2  \dd \Omega^2,
\label{Schwarzschild_metric_classic}
\end{equation}
with
\begin{subequations}
\begin{align} 
\mathcal{B}(r) &= 1-\dfrac{2GM}{c^2r}, \label{B(r)_classic}
\\
\mathcal{A}(r) &= \left(1-\dfrac{2GM}{c^2r}\right)^{-1},
\label{A(r)_classic}
\\
{\rm d}\Omega^2 &= {\rm d}\theta^2 + \sin^2 \theta \; {\rm d}\phi^2.
\end{align}
\end{subequations}
The radial coordinate $r$ has a simple geometric interpretation, as  it is related to the (proper) area $\mathscr{A}$ of the two-dimensional sphere with fixed $r$ and $t$ by the standard formula
\begin{equation}
r= \sqrt{\dfrac{\mathscr{A}}{4 \pi}}.
\end{equation} 

Another useful set of coordinates widely employed in the literature is represented by the harmonic coordinates $X^\mu=(X^0,X^1,X^2,X^3)$,  which can be introduced via the transformation \cite{Weinberg1972,Asanov1987}
\begin{align} \label{harmonic_coordinates_classic}
\left \{ \begin{array}{rl}
& X^0=ct,\\
& X^1= \mathcal{R}(r) \sin \theta \cos \phi, \\
& X^2= \mathcal{R}(r) \sin \theta \sin \phi, \\
& X^3= \mathcal{R}(r) \cos \theta, \\
\end{array}
\right.
\end{align} 
where 
\begin{align}
\mathcal{R}(r):= \sqrt{\left(X^1\right)^2+\left(X^2\right)^2+\left(X^3\right)^2}. 
\label{mathcal-R-classic} 
\end{align} 
In this case,  the Schwarzschild metric assumes the  form
\begin{equation}  \label{Schwarzschild_metric_harmonic_1_classic}
{\rm d}s^2 = G_{\mu \nu} {\rm d}X^\mu {\rm d}X^\nu = - \mathcal{B}\left(\mathcal{R}\right) \left(c^2{\rm d}t^2\right) + \left[ \dfrac{r^2\left(\mathcal{R}\right)}{\mathcal{R}^2} \delta_{ij} + \dfrac{X_i X_j}{\mathcal{R}^2} \left(\dfrac{\mathcal{A}\left(\mathcal{R}\right)}{\left(\mathcal{R}^\prime\right)^2}-\dfrac{r^2\left(\mathcal{R}\right)}{\mathcal{R}^2}\right)\right] {\rm d}X^i {\rm d}X^j, 
\end{equation}
where a prime denotes differentiation with respect to $r$ variable (e.g. $\mathcal{R}^\prime \equiv {\rm d}\mathcal{R}(r)/{\rm d}r$) and the  correspondence with the Newtonian theory demands that
\begin{equation} \label{limit_classic}
\lim_{\mathcal{R}\to 	\infty} \dfrac{r\left(\mathcal{R}\right)}{\mathcal{R}}=1.
\end{equation}
 
The explicit expression of the function \eqref{mathcal-R-classic} can be obtained from the harmonic condition $ \Box X^\mu=0$, with
 \begin{equation} \label{box_X_classic}
 \Box X^\mu \equiv g^{\alpha\beta} \nabla_\alpha \nabla_\beta X^\mu=g^{\alpha \beta} \left(\dfrac{\partial^2 X^\mu}{\partial x^\alpha \partial x^\beta}-\Gamma^{\lambda}_{\alpha \beta}\dfrac{\partial X^\mu}{\partial x^\lambda}\right).
 \end{equation}
Starting from  Eq. (\ref{harmonic_coordinates_classic}), a straightforward calculation  gives
\bsubeqs
\label{Box-harm-classic}
\begin{align}
\Box X^0 &=0,
\\
\Box X^i &= \left(\dfrac{X^i}{\mathcal{A}(r)\mathcal{R}(r)} \right) \left[\mathcal{R}^{\prime \prime}(r)+\left(\dfrac{\mathcal{B}^\prime(r)}{2 \mathcal{B}(r)}+\dfrac{2}{r} -\dfrac{\mathcal{A}^\prime(r)}{2 \mathcal{A}(r)}\right)\mathcal{R}^\prime(r)  -\dfrac{2 \mathcal{A}(r)}{r^2} \mathcal{R}(r)\right], \label{box_X_i_classic}
\end{align}
\esubeqs
which shows that the coordinates $X^\mu=(X^0,X^1,X^2,X^3)$ are harmonic provided that the term enclosed  in the square brackets of Eq. (\ref{box_X_i_classic}) vanishes.  Bearing in mind Eqs. (\ref{B(r)_classic}) and (\ref{A(r)_classic}), this amounts to require that $\mathcal{R}(r)$ satisfies the 
ordinary differential equation
\begin{equation} \label{equation-of-R-classic}
\mathcal{R}^{\prime \prime}(r) \left[r \left( r -2\dfrac{GM}{c^2}\right) \right] + \mathcal{R}^\prime(r) \left[2 \left( r -\dfrac{GM}{c^2}\right) \right] - 2  \mathcal{R}(r)=0,
\end{equation}
which can be written equivalently as a Legendre differential equation \cite{Abramowitz1964}
\begin{equation} 
\dfrac{{\rm d}}{{\rm d}r}  \left[r^2 \left(1-2\dfrac{GM}{c^2r}\right) \mathcal{R}^\prime(r)\right] -2  \mathcal{R}(r)=0,
\label{Legengre_equation}
\end{equation}
whose solution reads as
\bsubeqs \label{sol_R_2_classic}
\begin{align} 
\mathcal{R}(r)&= \mathscr{C}_1 P_1\left(\dfrac{c^2r}{GM}-1\right) + \mathscr{C}_2 Q_1\left(\dfrac{c^2r}{GM}-1\right),
\label{sol_R_1_classic}
\\
P_1(x) &\equiv x,
\\
Q_1(x) &\equiv P_1(x) Q_0(x)-1 \equiv \dfrac{x}{2} \log \left(\dfrac{x+1}{x-1}\right)-1, 
\end{align}
\esubeqs
where $\mathscr{C}_1$ and $\mathscr{C}_2$ are dimensionful integration constants depending on the ratio $GM/c^2$, while $P_n(x)$ and $Q_n(x)$ denote the Legendre polynomials  and Legendre functions of second kind (of degree $n$), respectively \cite{Abramowitz1964}. 

It is possible to write  the function (\ref{sol_R_1_classic}) in terms of a dimensionless constant $\bar{\mathscr{C}}_1$ and a  dimensionful one $\bar{\mathscr{C}}_2$   which do not depend on $GM/c^2$. Following Ref. \cite{Asanov1987}, we first consider Eq. \eqref{Legengre_equation} for $M=0$ 
\begin{equation} 
\dfrac{{\rm d}}{{\rm d}r}  \left[r^2  \mathcal{R}^\prime(r)\right] -2  \mathcal{R}(r)=0,
\end{equation}
which is solved by 
\begin{align}
\mathcal{R}(r)=\bar{\mathscr{C}}_1 r +  \bar{\mathscr{C}}_2 \frac{1}{r^2}.
\label{R-r-with-zero-mass}
\end{align}
Then, we plug into the solution \eqref{sol_R_1_classic} the expressions of  $Q_1\left(\dfrac{c^2r}{GM}-1\right)$ and $P_1\left(\dfrac{c^2r}{GM}-1\right)$ in the limit in which  $GM/c^2r$ approaches zero \cite{Abramowitz1964}
\begin{align}
Q_1\left(\dfrac{c^2r}{GM}-1\right) &= \frac{1}{3} \left(\frac{GM}{c^2 r}\right)^2+ \frac{2}{3} \left(\frac{GM}{c^2 r}\right)^3 + \dots,
\nonumber \\
P_1\left(\dfrac{c^2r}{GM}-1\right) &\approx \dfrac{c^2r}{GM}.
\end{align}
In this way,  one obtains for small values of $GM/c^2r$ the leading-order result
\begin{align}
\mathcal{R}(r)&= \mathscr{C}_1 \dfrac{c^2r}{GM}  +  \frac{\mathscr{C}_2}{3} \left(\dfrac{GM}{c^2r}\right)^2,    
\end{align}
and hence a comparison with Eq. \eqref{R-r-with-zero-mass} yields
\begin{align}
\mathscr{C}_1 &= \frac{GM}{c^2}    \bar{\mathscr{C}}_1,
\nonumber \\
\mathscr{C}_2 &= 3\left(\frac{c^2}{GM}\right)^2  \bar{\mathscr{C}}_2,
\end{align} 
which permit to rewrite  Eq. (\ref{sol_R_1_classic}) as
\begin{equation}
\mathcal{R}(r)=\bar{\mathscr{C}}_1 P_1\left(r-\dfrac{GM}{c^2}\right) + 3 \bar{\mathscr{C}}_2 \left(\dfrac{c^2}{GM}\right)^2 Q_1\left(\dfrac{c^2r}{GM}-1\right).
\end{equation}
The sufficient and necessary condition which permits  to  satisfy the requirement (\ref{limit_classic}) corresponds to set in the above equation $\bar{\mathscr{C}}_1=1$, with $\bar{\mathscr{C}}_2$ remaining arbitrary; in particular, if we choose $\bar{\mathscr{C}}_2=0$,  we end  up with the well-known  solution \cite{Weinberg1972,Asanov1987}
\begin{equation}\label{sol_R_classic}
\mathcal{R}(r)= r-\dfrac{GM}{c^2}.
\end{equation} 
Therefore,  it follows from Eqs. (\ref{Schwarzschild_metric_harmonic_1_classic}) and (\ref{sol_R_classic}) that the Schwarzschild metric in harmonic coordinates can be written as
\begin{align}\label{Schwarzschild_metric_harmonic_classic}
{\rm d}s^2 &= G_{\mu \nu} {\rm d}X^\mu {\rm d}X^\nu= 
\nonumber \\
&-\left(\frac{1-\frac{GM}{c^2 \mathcal{R}}}{1+\frac{GM}{c^2 \mathcal{R}}}\right) \left(c^2{\rm d}t^2\right) + \left[ \left(1+\frac{GM}{c^2 \mathcal{R}}\right)^2 \delta_{ij} + \dfrac{X_i X_j}{\mathcal{R}^2} \left(\frac{1+\frac{GM}{c^2 \mathcal{R}}}{1-\frac{GM}{c^2 \mathcal{R}}} \right)\left(\frac{GM}{c^2 \mathcal{R}}\right)^2\right] {\rm d}X^i {\rm d}X^j.
\end{align}

\subsection{The quantum corrected Schwarzschild metric}
\label{Sec:Quantum-corrected-Schwarzschild-metric}

In this section,  we will describe the strategy which we have adopted to derive  the quantum corrected Schwarzschild metric in standard coordinates. 

It will  soon be clear that we will deal with a situation which differs from the one presented in the previous section. Indeed, in Sec. \ref{Sec:classical-case},  we have exploited the knowledge of the (classical)  metric in Schwarzschild coordinates (cf. Eq. (\ref{Schwarzschild_metric_classic})) to eventually derive its     harmonic-coordinate expression in Eq. (\ref{Schwarzschild_metric_harmonic_classic}). On the other hand, in this section we will address a reversed scenario, as we will  begin with the quantum  metric  in harmonic coordinates (see Eq. (\ref{quantum_Schwarzschild_harmonic-2}) below) and  we will work out its form in  standard ones (see Eq. \eqref{Schwarzschild_metric_quantum-standard} below).

For this purpose, let us start with the main result of  Ref. \cite{Bohr2003}, where it has been shown how EFT techniques permit to compute the one-loop classical and quantum corrections to the Schwarzschild metric. By employing harmonic coordinates $X^\mu=(X^0,X^1,X^2,X^3)$, the metric reads as
\begin{align}
G_{00}&=-\left[1-2\dfrac{GM}{c^2 R} + 2 \dfrac{G^2M^2}{c^4 R^2} +k_1 \dfrac{G^2M\hbar}{c^5 R^3}+{\rm O}\left(G^3\right)+{\rm O}\left(G^3\hbar\right) \right], 
\nonumber \\
G_{0i}&=0, 
\nonumber \\
G_{ij} &=\delta_{ij} \left[1+2\dfrac{GM}{c^2 R} +  \dfrac{G^2M^2}{c^4 R^2} +k_2 \dfrac{G^2M\hbar}{c^5 R^3}+{\rm O}\left(G^3\right)+{\rm O}\left(G^3\hbar\right) \right] 
\nonumber \\
& + \dfrac{X_i X_j}{R^2} \left[\dfrac{G^2 M^2}{c^4 R^2} + k_3 \dfrac{G^2M\hbar}{c^5 R^3} +{\rm O}\left(G^3\right)+{\rm O}\left(G^3\hbar\right)\right],
\label{quantum_Schwarzschild_harmonic}
\end{align}
where ${\rm O}\left(G^3\right)$   and ${\rm O}\left(G^3\hbar\right)$ refer to the  two-loop classical and (leading\footnote{At two-loop level, apart from the leading quantum corrections proportional to $G^3 \hbar$, we also expect on general grounds subleading terms proportional to $G^3 \hbar^2$.}) quantum contributions, respectively, and (cf. Eq. \eqref{mathcal-R-classic})
\begin{align}
R := \sqrt{\left(X^1\right)^2 +\left(X^2\right)^2+\left(X^3\right)^2}.
\end{align}
The dimensionless parameters $k_i$ ($i=1,2,3$) can be obtained either from a detailed application of the Feynman diagrams' techniques \cite{Bohr2003,Donoghue1994} or the modern on-shell unitarity-based methods \cite{Bern1994,Bern2002,Bern2011,Bjerrum-Bohr2013}. However, since it seems that  in the literature there is a discrepancy  on their actual value  (see e.g. Table 1 in Ref. \cite{Bargueno2016}),   we have decided to  keep them as general as possible. Inspired by  Ref. \cite{Bohr2003}, it is possible to employ the following conventions: 
\begin{equation}
k_1 = \alpha k, \quad k_2 = \beta k, \quad k_3 = \gamma k, \quad (\alpha,\beta,\gamma,k \in \mathbb{R}),
\label{constants-k}
\end{equation} 
where $k$ is defined as the constant appearing in the quantum component of the (one-particle reducible) potential  \citep{Bohr2003}
\begin{equation}
V(R)= -\dfrac{Gm_1m_2}{R} \left[1-\dfrac{G(m_1+m_2)}{c^2 R}-k \dfrac{G \hbar}{c^3 R^2} + {\rm O}(G^2) +{\rm O}\left(G^2\hbar\right) \right]. 
\label{quantum-coorected-potential}
\end{equation}
As an example,  the outcome of Ref. \cite{Bohr2003} leads to
\begin{equation}
\alpha= \dfrac{124}{167}, \quad \beta= \dfrac{28}{167}, \quad \gamma=\dfrac{152}{167}, \quad k=\dfrac{167}{30 \pi}. 
\label{numbers-bohr-paper}
\end{equation}

It is clear from Eq.  \eqref{quantum_Schwarzschild_harmonic} that  in the EFT framework loop diagrams can give rise to both classical and purely quantum contributions (see Ref. \cite{Holstein2004} for a thorough discussion). Despite that, in view of our forthcoming investigation, we pursue a different procedure and employ  what we might call a  \qm{hybrid} scheme. In this \qm{hybrid} approach, we  consider a form of the metric which presents, on the one hand,  the classical terms stemming from a \emph{full} loop calculation, and, on the other,  the quantum factors arising  from  the one-loop diagrams only. In other words, we write  the metric in such a way that it reproduces exactly the  classical part \eqref{Schwarzschild_metric_harmonic_classic} and  meanwhile  contains only the leading long-distance quantum corrections which can be read off from Eq. \eqref{quantum_Schwarzschild_harmonic}.  Therefore, the starting point of our analysis will be the following metric components written in harmonic coordinates:
\begin{align}
G_{00}&=-\left(\frac{1-\frac{GM}{c^2R}}{1+\frac{GM}{c^2R}}\right) -\left[k_1 \dfrac{G^2M\hbar}{c^5 R^3}+{\rm O}\left(R^{-4}\right) \right], 
\nonumber \\
G_{0i}&=0, 
\nonumber \\
G_{ij} &=\delta_{ij} \left(1+\dfrac{GM}{c^2 R}\right)^2 + \dfrac{X_i X_j}{R^2} \left(\frac{GM}{c^2 R}\right)^2 \left(\frac{1+\frac{GM}{c^2R}}{1-\frac{GM}{c^2R}}\right) 
\nonumber \\
&+\left[ k_2 \dfrac{G^2M\hbar}{c^5 R^3} \delta_{ij}  + \left(\dfrac{X_i X_j}{R^2} \right) k_3 \dfrac{G^2M\hbar}{c^5 R^3}+{\rm O}\left(R^{-4}\right) \right],
\label{quantum_Schwarzschild_harmonic-2}
\end{align}
where hereafter  two-loop quantum corrections are encoded in the remainder ${\rm O}\left(R^{-4}\right)$. Note that, consistently with the aforementioned  \qm{hybrid} pattern, Eq. \eqref{quantum_Schwarzschild_harmonic-2} reproduces the  exact classical result \eqref{Schwarzschild_metric_harmonic_classic} if $\hbar=0$.

In order to determine the metric in Schwarzschild coordinates, we first introduce a set of Schwarzschild-like coordinates $\bar{x}^\mu=(c\bar{t},\bar{r},\bar{\theta},\bar{\phi})$ which are related to the harmonic ones by the transformation
(cf. Eq. \eqref{harmonic_coordinates_classic})
\begin{align} \label{harmonic_coordinates_quantum}
\left \{ \begin{array}{rl}
& X^0=c \bar{t},\\
& X^1=  R(\bar{r}) \sin \bar{\theta} \cos \bar{\phi}, \\
& X^2= R(\bar{r}) \sin \bar{\theta} \sin \bar{\phi}, \\
& X^3= R(\bar{r}) \cos \bar{\theta}. \\
\end{array}
\right.
\end{align} 
It ties in with the philosophy of our \qm{hybrid} approach, that the radial variable $R(\bar{r})$  can be written as the sum of a classical term  $\mathcal{R}(\bar{r})$ mirroring  Eq. \eqref{sol_R_classic} and a quantum correction. Since we expect, by standard arguments, that the latter  receives one-loop contributions proportional to $G^2 M \hbar/\left(c^5\bar{r}^2\right)$,   the function $R(\bar{r})$  has the general form
\begin{align}\label{function-R-of-r-quantum}
R(\bar{r}) = \bar{r} -\frac{GM}{c^2} + \lambda \frac{G^2 M \hbar}{c^5 \bar{r}^2} + \OO\left(\bar{r}^{-3}\right),
\end{align} 
where $\lambda$ is a real-valued constant     and $ \OO\left(\bar{r}^{-3}\right)$ denotes two-loop quantum factors. The  inversion of Eq. \eqref{function-R-of-r-quantum} yields 
\begin{align}
\bar{r}(R)=R +\frac{GM}{c^2} - \lambda \frac{G^2 M \hbar}{c^5 R^2} + \OO\left(R^{-3}\right),
\end{align}
which means that the physical constraint (cf. Eq. \eqref{limit_classic})
\begin{align*}
\lim_{R \to \infty} \frac{\bar{r}(R)}{R}=1,
\end{align*}
is straightforwardly  satisfied.

We now need to find the value of the parameter $\lambda$. For this reason,   we use the fact that the metric $\bar{g}_{\alpha \beta}(\bar{x})$ in the quasi-standard coordinates $\bar{x}^\mu$  can be written via the usual relation
\begin{align} \label{metric-bar-coord}
\bar{g}_{\alpha \beta}(\bar{x}) = \frac{\partial X^\mu}{\partial \bar{x}^\alpha}    \frac{\partial X^\nu}{\partial \bar{x}^\beta} G_{\mu \nu},    
\end{align}
which, after a lengthy but straightforward calculation involving Eqs. \eqref{quantum_Schwarzschild_harmonic-2}--\eqref{function-R-of-r-quantum}, leads to
\begin{align}
\dd s^2 = \bar{g}_{\alpha \beta} {\rm d}\bar{x}^\alpha  {\rm d}\bar{x}^\beta=  -\bar{B}(\bar{r})  \left(c^2{\rm d}\bar{t}^2\right)+\bar{A}(\bar{r}) \dd \bar{r}^2 + \bar{C}(\bar{r}) \left(\dd \bar{\theta}^2 + \sin^2 \bar{\theta} \; \dd \bar{\phi}^2\right),
\end{align}
where
\begin{align}
\bar{B}(\bar{r}) &= \left(1-\frac{2GM}{c^2 \bar{r}}\right) + \left[k_1 \frac{G^2 M \hbar}{c^5 \bar{r}^3} + \OO\left(\bar{r}^{-4}\right)\right],   
 \nonumber\\
\bar{A}(\bar{r}) &= \left(1-\frac{2GM}{c^2 \bar{r}}\right)^{-1}  + \left[\left(k_2+k_3+2 \lambda\right) \frac{G^2 M \hbar}{c^5 \bar{r}^3} + \OO\left(\bar{r}^{-4}\right)\right],  
 \nonumber \\
 \bar{C}(\bar{r}) &= \bar{r}^2  + \bar{r}^2  \left[\left(k_2+2 \lambda\right) \frac{G^2 M \hbar}{c^5 \bar{r}^3} + \OO\left(\bar{r}^{-4}\right)\right].
 \label{variables-B,A,C-bar}
\end{align}
It is  simple to show that the coordinates \eqref{harmonic_coordinates_quantum} are harmonic (i.e.,  they fulfill the condition $\Box X^\mu =0$) provided that the radial variable \eqref{function-R-of-r-quantum}  satisfies the following differential equation (cf. Eq. \eqref{Box-harm-classic}):
\begin{align}\label{R-bar-diff-equation}
R^\prime \left(\frac{\bar{B}^\prime}{2\bar{B}}+\frac{\bar{C}^\prime}{\bar{C}}-\frac{\bar{A}^\prime}{2\bar{A}}\right) + R^{\prime \prime} - \frac{2\bar{A}}{\bar{C}} R=0,
\end{align}
where now a prime denotes a differentiation with respect to $\bar{r}$. It then follows from  formulas \eqref{variables-B,A,C-bar} that the above equation is satisfied modulo  $\OO\left(\bar{r}^{-5}\right)$ corrections (note in fact that the presence of the second-order derivative $R^{\prime \prime}$ implies that Eq. \eqref{R-bar-diff-equation} has to be worked out up to $\OO\left(\bar{r}^{-5}\right)$ terms, see Eq. \eqref{function-R-of-r-quantum}) if
\begin{align} \label{lambda-value}
    \lambda= -\frac{1}{2}\left(k_1+k_2+\frac{k_3}{3}\right).
\end{align}
Due to the last identity, the functions \eqref{variables-B,A,C-bar} depend  on  $k_1$ and $k_3$ only. We can further simplify the form assumed by the metric by introducing Schwarzschild coordinates
$x^\mu=(ct,r,\theta,\phi)$ via the following relations:
\begin{align} 
\left \{ \begin{array}{rl}
& t= \bar{t},\\
& r=  \sqrt{\bar{C}(\bar{r})}, \\
& \theta= \bar{\theta}, \\
& \phi= \bar{\phi}. \\
\end{array}
\right.
\end{align} 
In this way,  we obtain the sought-after form of the Schwarzschild metric in standard coordinates including one-loop quantum corrections
\begin{equation}
{\rm d}s^2=g_{\mu \nu}{\rm d}x^\mu{\rm d}x^\nu= - B(r)\left(c^2{\rm d}t^2\right) + A(r){\rm d}r^2 + r^2 {\rm d}\Omega^2,
\label{Schwarzschild_metric_quantum-standard}
\end{equation}
where
\bsubeqs 
\label{A-and-B-of-r-quantum}
\begin{align} 
B(r) &= 1-\dfrac{2GM}{c^2r} + \left[k_1 \frac{G^2 M \hbar}{c^5 r^3} + \OO\left(r^{-4}\right)\right],
\label{B-of-r-quantum}
\\
A(r) &= \left(1-\dfrac{2GM}{c^2r}\right)^{-1}+\left[-3k_1 \frac{G^2 M \hbar}{c^5 r^3} + \OO\left(r^{-4}\right)\right].
\label{A-of-r-quantum}
\end{align}
\esubeqs
As pointed out before, our metric is such that  $-g_{tt} \neq g^{rr}$.

For simplicity, the terms $\OO\left(r^{-4}\right)$ indicating two-loop quantum contributions will be hereafter omitted.

\section{Applications} \label{Sec:applications} 

In the last section, we have obtained the Schwarzschild metric involving the leading low-energy quantum corrections stemming from one-loop Feynman diagrams. The final result has been  expressed in Schwarzschild coordinates and can be found in  Eqs. \eqref{Schwarzschild_metric_quantum-standard} and \eqref{A-and-B-of-r-quantum}. In this section, we explore some features of this  quantum geometry. The study of the horizon(s) is given in Sec. \ref{Sec:horizon}, while the  behaviour of the metric component $g^{rr}$, which reveals the presence of null hypersurfaces having constant radius,  is investigated in Sec. \ref{Sec:analysis-g-rr}. Then, we work out the curvature invariants in Sec. \ref{Sec:curvature-inv} and conclude the section with a first analysis of the geodesic motion (see Sec. \ref{Sec:analysis-geod}). 

In our forthcoming investigation, we will suppose that 
\begin{align} \label{Rs-bigger-than-lp}
  R_{\rm S} \gg \ell_{\rm P},  
\end{align}
or, equivalently,
\begin{align} \label{M-bigger-M-Planck}
  M \gg \frac{1}{2}  M_{\rm P},
\end{align}
$R_{\rm S} = 2GM/c^2$, $\ell_{\rm P}=\left(G \hbar/c^3\right)^{1/2}$, and $M_{\rm P} = \left(\hbar c/G\right)^{1/2}$   being  the Schwarzschild radius,   Planck length, and  Planck mass, respectively.  Furthermore, we will assume that these conditions are not spoiled if both sides are multiplied by $\vert k_1 \vert$, since  the typical values reported in the literature are such that  $ \vert k_1 \vert  \sim {\rm O} (1)$ (see e.g. Ref. \cite{Bargueno2016}). 

As will be clear from next sections, the above relations permit  to write our results as a classical piece corrected by a  small  quantum factor. Furthermore, they imply that our examination excludes the so-called primordial or micro black holes (see Refs. \cite{Carr1975,Niemeyer1999,Green2020,Carr2021}, for further details), i.e., objects having a mass of the order of  $M_{\rm P}$ and for which quantum  phenomena assume a crucial role. In this way, we do not enter a full quantum-gravity regime where the  EFT scheme breaks down.

\subsection{Metric horizons } \label{Sec:horizon}

Since the metric \eqref{Schwarzschild_metric_quantum-standard} pertains to a static, asymptotically-flat spacetime, the horizon(s) can be identified by the equation $g_{tt}=0$ \cite{Morris-Thorne1988}. This condition, with the help of Eq. \eqref{B-of-r-quantum}, leads to the following cubic equation:
\begin{equation}\label{cubic-horizon}
    r^3 -\RS r^2 + \frac{1}{2}k_1 \RS \LP^2=0,
 \end{equation}
whose discriminant reads as
\begin{align}\label{Delta-cubic}
    \Delta=2 k_1 R^4_{\rm S} \ell_{\rm P}^2\left(1 -\frac{27}{8} k_1 \frac{\ell_{\rm P}^2}{R^2_{\rm S}} \right).
\end{align}
The nature of the roots of the cubic \eqref{cubic-horizon} are determined by  the sign of $\Delta$\footnote{Recall that an algebraic equation of third degree with real coefficients  admits three distinct real roots if $\Delta>0$, multiple real roots when $\Delta=0$, and  one real root jointly with two complex conjugate roots if $\Delta <0$.\label{footnote-cubic}}, which in turn depends on that of $k_1$. If $k_1$ is negative, then $\Delta <0$ for any real-valued $M$. Thanks to Descartes' rule of signs, we see that Eq. \eqref{cubic-horizon} admits one positive real solution and two complex conjugate solutions. On the other hand, when $k_1$ attains positive values  the discriminant \eqref{Delta-cubic} can be either positive,  negative,  or vanishing. If $M<M^\star$, with
\begin{align}
M^\star :=  M_{\rm P} \, \sqrt{\frac{27}{32}k_1},
\label{M-star}    
\end{align}
then $\Delta<0$, whereas for $M \geq M^\star$ we have $\Delta \geq 0$. In the first case, the cubic \eqref{cubic-horizon} admits one real negative root and two complex conjugate roots, while in the second it provides two positive real solutions (which coincide if $\Delta=0$) and one real negative solution. Therefore, to sum up, the algebraic equation \eqref{cubic-horizon} is such that
\begin{align}
k_1<0 \Rightarrow \Delta<0 \;\, \forall M \in \mathbb{R}: \quad  \mbox{one positive root and two complex roots},   
\label{case-k1-negative}
\end{align}
\begin{align}
k_1>0 \Rightarrow   \left \{ 
\setlength{\tabcolsep}{10pt} 
\renewcommand{\arraystretch}{1.8}
\begin{array}{rl}
& \Delta <0 \;\;  {\rm if} \; M<M^\star: \quad \mbox{one negative root and two complex roots},\\
& \Delta \geq 0 \;\;  {\rm if} \; M \geq M^\star:\quad \mbox{one negative root and   two positive roots}.
\end{array}
\right.
\label{case-k1-positive}
\end{align}

It is thus clear that the case with $k_1>0$ and $\Delta <0$ yields no physical solution. On the other hand, when both $k_1$ and $\Delta$ are positive, the two positive roots of Eq.  \eqref{cubic-horizon} can be conveniently parametrized in the trigonometric form. In this way,  we have
\begin{align}
r_1 &= \frac{2}{3}R_{\rm S} \cos \left\{ \frac{1}{3} \arccos \left[1-\frac{27}{4}k_1 \left(\frac{\ell_{\rm P}}{R_{\rm S}}\right)^2\right] \right\}   +\frac{1}{3} R_{\rm S} < \RS,
\label{solution-r-1}  
\\
r_2 &= \frac{2}{3}R_{\rm S} \cos \left\{ \frac{1}{3} \arccos \left[1-\frac{27}{4}k_1 \left(\frac{\ell_{\rm P}}{R_{\rm S}}\right)^2\right]-\frac{2 \pi}{3} \right\}   +\frac{1}{3} R_{\rm S}
 \nonumber \\
&=-\frac{2}{3}R_{\rm S} \sin \left\{\frac{\pi}{6}- \frac{1}{3} \arccos \left[1-\frac{27}{4}k_1 \left(\frac{\ell_{\rm P}}{R_{\rm S}}\right)^2\right] \right\}   +\frac{1}{3} R_{\rm S},
\label{solution-r-2}
\end{align}
which, owing to Eq. \eqref{Rs-bigger-than-lp}, can be approximated as
\begin{align}
r_1 &= R_{\rm S} \left[1-\frac{k_1}{2} \frac{\LP^2}{\RS^2} + \OO\left(\ell_{\rm P}^4/R_{\rm S}^4\right)\right],
\label{solution-r-1-expanded}  
\\
r_2 &= \ell_{\rm P} \sqrt{\frac{k_1}{2}} + \OO\left(\ell_{\rm P}^2/R_{\rm S}\right).
\label{second-horizon}
  \end{align}
It is worth observing  that, in view of Eq. \eqref{Rs-bigger-than-lp} jointly with the fact that $\vert k_1 \vert \sim {\rm O}(1)$, we can assume  that
\begin{align}
r_2 \ll r_1 \label{r2-smaller-r1}.
\end{align}
Moreover, we note that Eqs. \eqref{solution-r-1} and \eqref{solution-r-2} are well-defined functions since we are supposing $M>M^\star$, which is the condition that implies $\Delta>0$ when  $k_1>0$ (cf. Eqs. \eqref{M-star} and \eqref{case-k1-positive}). 

As pointed out before, the discriminant \eqref{Delta-cubic} vanishes  when $M=M^\star$. In this case, Eqs.  \eqref{solution-r-1} and \eqref{solution-r-2} boil down to the single root
\begin{align} \label{r-star}
r^\star =  \ell_{\rm P} \sqrt{\frac{3}{2}k_1}.   
\end{align}

When $k_1<0$ (and hence $\Delta<0$), the only positive solution of the cubic  \eqref{cubic-horizon}  can be easily written through the Cardano formula, which yields
\begin{align}
r_3 &= \frac{1}{3} R_{\rm S} + \frac{1}{3}  R_{\rm S} \left[1-\frac{27}{4}k_1 \left(\frac{\ell_{\rm P}}{R_{\rm S}}\right)^2 + 3 \sqrt{3} \, \sqrt{\frac{k_1}{2}\left(\frac{\ell_{\rm P}}{R_{\rm S}}\right)^2 \left(k_1\frac{27}{8} \frac{\ell_{\rm P}^2}{R_{\rm S}^2}-1\right)}\right]^{-1/3}  
\nonumber \\
&+ \frac{1}{3}  R_{\rm S} \left[1-\frac{27}{4}k_1 \left(\frac{\ell_{\rm P}}{R_{\rm S}}\right)^2 + 3 \sqrt{3} \, \sqrt{\frac{k_1}{2}\left(\frac{\ell_{\rm P}}{R_{\rm S}}\right)^2 \left(k_1\frac{27}{8} \frac{\ell_{\rm P}^2}{R_{\rm S}^2}-1\right)}\right]^{1/3}  
\nonumber \\
&= R_{\rm S} \left[ 1-\frac{k_1}{2} \frac{\LP^2}{\RS^2} +\OO\left(\ell_{\rm P}^4/R_{\rm S}^4\right) \right]  > \RS,
\label{quantum-horizon}
\end{align}
where in the last equality  we have exploited the relation \eqref{Rs-bigger-than-lp} to  neglect higher-order corrections.

At this point, some comments are in order. First of all,  we notice that the radii $r_1$ and $r_3$   assume the same form modulo higher-order corrections (in fact, they only differ by the sign of $k_1$, see Eqs. \eqref{solution-r-1-expanded} and \eqref{quantum-horizon}).  Moreover, our investigation proves that, when $k_1<0$,  the cubic   \eqref{cubic-horizon} presents one positive root. Although this result is valid, in principle, for any   mass $M$,  it amounts to the Schwarzschild radius plus tiny quantum corrections provided that Eqs. \eqref{Rs-bigger-than-lp} or \eqref{M-bigger-M-Planck} are employed. On the other hand, the scenario with both  $k_1$ and $\Delta$   positive  predicts the existence of two horizons when $M>M^\star$, a condition which is consistent with  Eq. \eqref{M-bigger-M-Planck}. This second case permits to appreciate the  importance of  the relations \eqref{Rs-bigger-than-lp} and \eqref{M-bigger-M-Planck}. Indeed, without these constraints, masses smaller than $M^\star$ could  be taken into account and hence we would end up with a spacetime possessing no horizon (cf. Eq. \eqref{case-k1-positive}); however, this situation seems unrealistic, since we expect that  quantum theory leads to tiny departures from the classical Schwarzschild geometry in the low-energy regime. Therefore, Eqs. \eqref{Rs-bigger-than-lp} and \eqref{M-bigger-M-Planck} are crucial, since, as pointed out before, they allow us to put physical results in the form of standard classical quantities modified by some small quantum contributions. 

The framework with $k_1>0$ exhibits one drawback. In fact, there exist values of $k_1$ for which the   horizon \eqref{second-horizon} can have a radius smaller than $\ell_{\rm P}$\footnote{The same conclusion is true also for the solution \eqref{r-star}, which is valid when $M=M^\star$. However, this case can be discarded owing to Eq. \eqref{M-bigger-M-Planck}.}. This might be nonsensical, as it is believed that the current laws of physics should lose their validity below the Planck scale. Indeed, the  Planck length is thought to be the smallest possible length, as the known principles of quantum mechanics and gravity indicate that it is impossible  to measure the position of an object with a precision smaller than $\ell_{\rm P}$ \cite{Mead1964}. Furthermore, it has been proved that $\ell_{\rm P}$ represents a lower bound to all proper length scales \cite{Padmanabhan1985}. Despite these shortcomings, the scenario having positive $k_1$ can still have a physical meaning thanks to the inequality \eqref{r2-smaller-r1}, which we recall follows from the condition \eqref{Rs-bigger-than-lp}. In fact, as we will show in the next section, close to the horizon located at $r=r_1$ there exists the null hypersurface $r=\tilde{r}_1=\RS + \OO \left(\LP^2 / \RS\right)$ which separates regions of the spacetime where $r=const$ is a  timelike hypersurface  from those where $r=const$ corresponds to a spacelike hypersurface (see Eq. \eqref{solution-r-tilde-1} below). Since it is hidden by the null hypersurface $r=\tilde{r}_1$, the  horizon at $r=r_2$ is not visible to any observer lying in the domain $r > \RS + \OO \left(\LP^2 / \RS\right)$, making  the potential relation $r_2 < \LP$  harmless.

\subsection{Analysis of the  radial component $g^{rr}$}\label{Sec:analysis-g-rr}

As we have noted before, the quantum metric \eqref{Schwarzschild_metric_quantum-standard} has the peculiar property that the components $-g_{tt}$ and $g^{rr}$ differ. For this reason, the condition $g_{tt}=0$ does not define the locus where the hypersurface $r= const $ becomes null. Thus, this analysis will be performed in this section. Moreover, a possible energy-extraction phenomenon is discussed in Sec. \ref{Sec:energy-extraction}.  

Let us recall that the normal $n^\mu$ to the  hypersurface $r= const $ satisfies
\begin{align}
n_\mu n^\mu =g^{rr},
\label{n-mu-n-mu}
\end{align}
and (cf. Eq. \eqref{A-of-r-quantum}) 
\begin{align} \label{g-rr-inv-quatum}
g^{rr}=  1-\frac{\RS}{r} + \frac{3}{2}k_1 \frac{\RS \LP^2}{r^3}, 
\end{align}
where, consistently with the EFT scheme, we have neglected  $\OO\left(r^{-4}\right)$ corrections.  The condition $g^{rr}=0$ leads to the following cubic equation:
\begin{align}
r^3 - r^2 \RS + \frac{3}{2} k_1 \RS \LP^2=0,   
\label{cubic-null-hypersurface}
\end{align}
which   can be investigated via the same procedure as in Sec. \ref{Sec:horizon}. Therefore,  when $k_1$ is negative, the discriminant of the cubic \eqref{cubic-null-hypersurface} is negative for any real-valued $M$; on the other hand, if $k_1 >0$ the discriminant is negative (resp. positive) provided that $M< \sqrt{3} M^\star$ (resp. $M> \sqrt{3} M^\star$). 
In the scenario $k_1>0$  and $M> \sqrt{3} M^\star$, the two real positive roots of the cubic \eqref{cubic-null-hypersurface} read as
\begin{align}
\tilde{r}_1 &= \frac{2}{3}R_{\rm S} \cos \left\{ \frac{1}{3} \arccos \left[1-\frac{81}{4}k_1 \left(\frac{\ell_{\rm P}}{R_{\rm S}}\right)^2\right] \right\}   +\frac{1}{3} R_{\rm S} 
 \nonumber \\
&= R_{\rm S} \left[1-\frac{3 k_1}{2} \frac{\LP^2}{\RS^2} + \OO\left(\ell_{\rm P}^4/R_{\rm S}^4\right)\right],
\label{solution-r-tilde-1}  
 \\
\tilde{r}_2 &= -\frac{2}{3}R_{\rm S} \sin \left\{\frac{\pi}{6}- \frac{1}{3} \arccos \left[1-\frac{81}{4}k_1 \left(\frac{\ell_{\rm P}}{R_{\rm S}}\right)^2\right] \right\}   +\frac{1}{3} R_{\rm S}
\nonumber \\
&= \LP \sqrt{\frac{3k_1}{2}} + \OO\left(\LP^2/\RS\right),
\label{solution-r-tilde-2}
\end{align}
and in our hypotheses (recall Eq. \eqref{Rs-bigger-than-lp} and the fact that $\vert k_1 \vert \sim {\rm O} \left(1\right)$) satisfy 
\begin{align}
\tilde{r}_2 \ll \tilde{r}_1; 
\end{align}
moreover, a comparison with Eqs. \eqref{solution-r-1} and \eqref{solution-r-2} (or equivalently with Eqs. \eqref{solution-r-1-expanded} and \eqref{second-horizon}) shows that
\begin{align}
\tilde{r}_1 &< r_1 < \RS,
\label{inequality-r1-tilde-r1}
\\
\tilde{r}_2 & > r_2.
\end{align}
\begin{figure}[bht!]
\centering\includegraphics[scale=0.30]{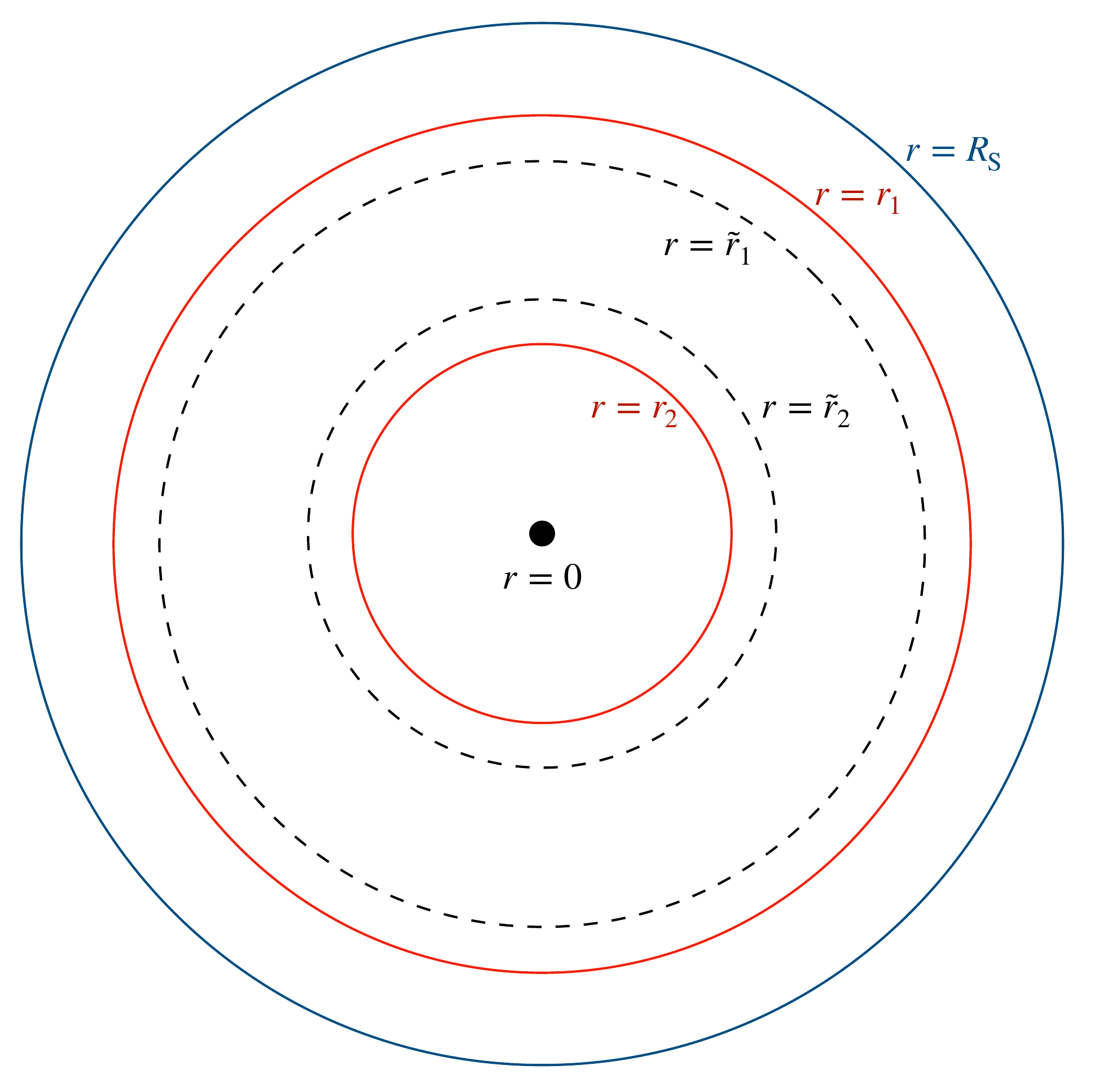}
\caption{Pictorial representation of the  Schwarzschild radius,  horizons located at $r=r_1$ and $r=r_2$, and null hypersurfaces  $r=\tilde{r}_1$ and $r=\tilde{r}_2$ in the case $k_1 >0$. }
\label{Fig-k1-positive}
\end{figure}
The above inequalities imply that both the null hypersurface $r=\tilde{r}_2$ and the horizon at $r=r_2$ are hidden by the null hypersurface $r=\tilde{r}_1$ (see Fig. \ref{Fig-k1-positive}). Therefore, the value of $\tilde{r}_2$ and $r_2$, which might be smaller than the Planck length,  cannot be measured by any observer  located in the region outside the null hypersurface $r=\tilde{r}_1$.

\begin{figure}[bht!]
\centering\includegraphics[scale=0.35]{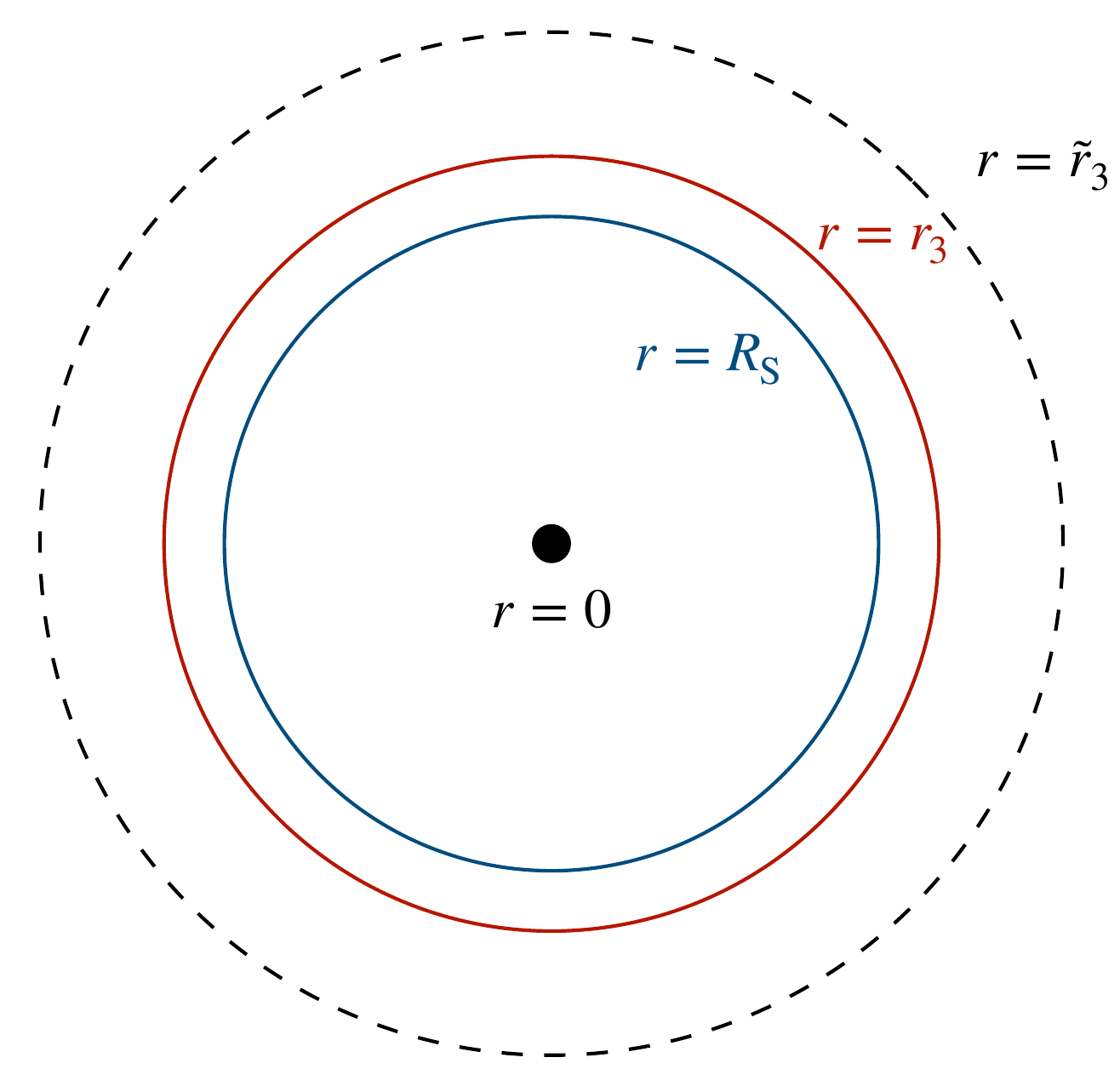}
\caption{Pictorial representation of  the null hypersurface  $r=\tilde{r}_3$, horizon located at $r=r_3$,  and  Schwarzschild radius in the case $k_1 <0$. }
\label{Fig-k1-negative}
\end{figure}

In the case with a negative $k_1$, the positive root of the cubic \eqref{cubic-null-hypersurface} is
\begin{align}
\tilde{r}_3 &= \frac{1}{3} R_{\rm S} + \frac{1}{3}  R_{\rm S} \left[1-\frac{81}{4}k_1 \left(\frac{\ell_{\rm P}}{R_{\rm S}}\right)^2 + 3 \sqrt{3} \, \sqrt{\frac{3k_1}{2}\left(\frac{\ell_{\rm P}}{R_{\rm S}}\right)^2 \left(k_1\frac{81}{8} \frac{\ell_{\rm P}^2}{R_{\rm S}^2}-1\right)}\right]^{-1/3}  \nonumber \\
&+ \frac{1}{3}  R_{\rm S} \left[1-\frac{81}{4}k_1 \left(\frac{\ell_{\rm P}}{R_{\rm S}}\right)^2 + 3 \sqrt{3} \, \sqrt{\frac{3k_1}{2}\left(\frac{\ell_{\rm P}}{R_{\rm S}}\right)^2 \left(k_1\frac{81}{8} \frac{\ell_{\rm P}^2}{R_{\rm S}^2}-1\right)}\right]^{1/3}  
\nonumber \\
&= R_{\rm S} \left[ 1-\frac{3k_1}{2} \frac{\LP^2}{\RS^2} +\OO\left(\ell_{\rm P}^4/R_{\rm S}^4\right) \right],
\label{solution-r-tilde-3}
\end{align}
and hence,  as a consequence of Eq. \eqref{quantum-horizon}, we have
\begin{align}
\tilde{r_3} >  r_3  > \RS,
\label{inequality-tilde-r3-r3}    
\end{align}
see Fig. \ref{Fig-k1-negative}.

From the above analysis it is clear that the null hypersurface(s) having constant radius and   the horizon(s) studied in Sec. \ref{Sec:horizon} do not coincide, as anticipated  before. Furthermore, we note that  upon neglecting $\OO\left(\ell_{\rm P}^4/R_{\rm S}^3\right)$ terms the radii $\tilde{r}_1$ and $\tilde{r}_3$  assume the same form. 

At this stage, let us examine the sign of $g^{rr}$. To give a  detailed analysis of its behaviour, we rely on the study  of the derivative $g^{\prime  rr}$ with respect to the radial variable
\begin{align} \label{grr-derivative}
g^{\prime  rr}= \frac{\RS}{r^4} \left(r^2 -\frac{9}{2} k_1 \LP^2 \right), 
\end{align}
and the knowledge of the limits
\begin{subequations}
\label{limits-grr}
\begin{align}
\lim_{r\to 0^+} g^{rr} &= \pm \infty,
\label{limit-grr-1}
 \\
\lim_{r\to +\infty} g^{rr} &=    1,
\end{align}
\end{subequations}
where in Eq. \eqref{limit-grr-1} the upper sign refers to the case $k_1>0$, while the lower one to the situation $k_1<0$. Therefore,  when $k_1$ is positive $g^{rr}$ attains a minimum at $r= 3 \LP \sqrt{k_1/2}$ and hence it follows from Eq. \eqref{limits-grr} that
\begin{subequations}
\begin{empheq}[left={k_1>0 \Rightarrow \empheqlbrace\,}]{align}
&  r < \tilde{r}_2: \;\, \; \quad \quad \qquad g^{rr}>0, \label{sign-grr-1} \\
& \tilde{r}_2 < r < \tilde{r}_1: \qquad \;  \, g^{rr} < 0, \\
&  r >\tilde{r}_1: \qquad \qquad \, \, \; g^{rr} > 0, 
\end{empheq}
\end{subequations}
which implies that (cf. Eq. \eqref{n-mu-n-mu}) all hypersurfaces  having a constant radius smaller than $\tilde{r}_2$ or larger than $\tilde{r_1}$ are timelike and hence can be crossed by a particle either inwards or outwards \cite{Ferrari2020}; in particular, the horizon at $r=r_1$ is a timelike hypersurface owing to Eq. \eqref{inequality-r1-tilde-r1}.  On the other hand, the hypersurfaces having a constant radius lying in the interval $ \tilde{r}_2 <r<\tilde{r}_1$ are spacelike, which means that they can be crossed by a particle in one direction only \cite{Ferrari2020}.

In the case $k_1 <0$, Eq. \eqref{grr-derivative} reveals that $g^{rr}$ is a monotonically increasing function with no stationary points. Therefore, bearing in mind Eq. \eqref{limits-grr}, we can claim that \begin{subequations}
\begin{empheq}[left={k_1<0 \Rightarrow \empheqlbrace\,}]{align}
&  r < \tilde{r}_3: \;\; \qquad \quad \;\;\, g^{rr}<0,  \label{sign-grr-4}\\
&  r >\tilde{r}_3: \qquad \qquad \, \, g^{rr} > 0, \label{sign-grr-5} 
\end{empheq}
\end{subequations}
which shows that   the hypersurfaces $r= const $ are spacelike if $r <\tilde{r}_3$, while  they are timelike otherwise; this  also means that  the horizon at $r=r_3$ is a spacelike hypersurface (see Eq. \eqref{inequality-tilde-r3-r3}). 

When $k_1$ is positive, an intriguing comparison with the classical (nonextreme) Reissner-Nordstr\"{o}m geometry can be made. It is known  \cite{Carroll2004,Poisson2009} that the Reissner-Nordstr\"{o}m metric represents a solution of Einstein-Maxwell equations describing the gravitational field of a static, spherically symmetric, and electrically charged source. The most interesting features of this pattern emerge when a black hole is taken into account, as it possesses two horizons, the inner one being  an apparent horizon and the outer one  an event horizon. In this setup, the singularity located at $r=0$ represents a timelike hypersurface which can be avoided by observers moving within the black hole. In particular, after having crossed the outer and the inner horizons, the observer can traverse (another copy of) the latter  again,  emerge out of the black hole, and then enter a new external and asymptotically-flat universe.

Something similar might occur in the quantum Schwarzschild spacetime in the framework $k_1 >0$, where $r=0$ is a timelike hypersurface (cf. Eq. \eqref{sign-grr-1}, while for $k_1 <0$ it is a spacelike hypersurface, see Eq. \eqref{sign-grr-4}). Indeed, in this scenario, the observer going through the hypersurface $r=\tilde{r}_2$ can decide to reverse his/her journey at any moment (i.e., either before or after having travelled through the horizon located at $r=r_2$), thus evading  the singularity at $r=0$. However, a precise  assessment of this situation requires the examination of the maximal extensions of the quantum metric \eqref{Schwarzschild_metric_quantum-standard}. This is a delicate problem which is  more complicated than in the case of the classical Schwarzschild geometry due to  the fact that in the quantum regime the components $-g_{tt}$ and $g^{rr}$ differ. For this reason, this point deserves a careful investigation in a separate paper.

\subsubsection{A possible energy-extraction process}\label{Sec:energy-extraction}

The horizon located at $r=r_1$ is such that  $r_1>\tilde{r}_1$, see Eq. \eqref{inequality-r1-tilde-r1} and Fig. \ref{Fig-k1-positive}. Thus, as pointed out in the previous section, it can be traversed either inwards or outwards, since $g^{rr}$ attains positive values at  $r=r_1$. The same conclusions hold also for the hypersurfaces  having constant radius which lies  in the interval $\tilde{r}_1 < r <r_1$. Therefore, when $k_1$ is positive a peculiar phenomenon can occur in the region $\tilde{r}_1 < r <r_1$, as we are going to show. 

First of all, we need to  analyze the sign of the temporal component $g_{tt}$ of the metric tensor (see Eq. \eqref{B-of-r-quantum}). Similarly as before, this task can be fulfilled in the following way. The derivative $g^\prime_{tt}$ with respect to the $r$ variable reads as
\begin{align}
 g^\prime_{tt}= \frac{R_{\rm S}}{2 r^4}  \left(3 k_1 \ell_{\rm P}^2 -2 r^2\right); 
\end{align}
furthermore, owing to the asymptotic flatness property, we know that $\lim \limits_{r \to +\infty} g_{tt} = -1$; moreover,  $g_{tt}$ satisfies the following condition: 
\begin{align}
& \lim \limits_{r \to 0^+} g_{tt} = \mp \infty,
\end{align}
the upper (resp. lower) sign referring to the case $k_1 > 0$ (resp. $k_1 < 0$). The above relations permit to conclude that, when $k_1 $ is positive, $g_{tt}$ attains a maximum at $r=r^\star$ (cf. Eq. \eqref{r-star}) and hence it  is negative when $r<r_2$ or $r>r_1$, while it is positive otherwise. This means, in particular,   that  the static Killing vector field $K^\mu=(1,0,0,0)$ becomes spacelike if $\tilde{r}_1 < r <r_1$. This fact entails a twofold consequence. Firstly, no static observer can exist
in  the domain $\tilde{r}_1 < r <r_1$; secondly, the (conserved) energy  $\mathscr{E}=-c K_\mu P^\mu$ of a freely falling particle having four-momentum $P^\mu$ can assume negative values as soon as its  radial coordinate satisfies $\tilde{r}_1 < r <r_1$. Therefore, the region $\tilde{r}_1 < r <r_1$ has all the features needed to conceive an energy-extraction mechanism similar to the Penrose process occurring in Kerr spacetime \cite{Wald}.  However, two comments are in order: $(i)$ the   region $\tilde{r}_1 < r <r_1$ has an extension of the order of the ratio $\ell_{\rm P}^2/R_{\rm S}$ (cf. Eqs. \eqref{solution-r-1-expanded} and \eqref{solution-r-tilde-1}); $(ii)$  no extraction of rotational energy is involved in this phenomenon. Most importantly, the process we are considering  violates the Hawking area theorem (or, equivalently, the second law of black-hole mechanics), since it causes  the mass of the black hole to be reduced to $M - c^{-2} \vert \mathscr{E} \vert$, which means that the corresponding  variation   $\delta M$  is negative. In addition,  $\delta M$ is not bounded and hence there seems to be  no limit on the energy that can be extracted from the black hole, which would thus undergo an inevitable  evaporation phase. 

Hawking area theorem holds provided that the null energy condition (NEC) is enforced \cite{Wald,Wald-LRR}. This means that the proposed phenomenon could be valid if the quantum Schwarzschild  metric is  sourced by a stress-energy tensor which does not respect the NEC (examples where the NEC is violated are furnished by the case of exotic matter or quantum fields, see e.g. Refs. \cite{Morris-Thorne1988,Rubakov2014,DiGrezia2017,Kontou2020} for further details). This conclusion seems to support the examination of  Ref. \cite{Bohr2003}, where it is proved that the quantum metric \eqref{quantum_Schwarzschild_harmonic-2} can be derived from the source stress-energy tensor, which, because of the radiative corrections affecting its form,  contains the contributions of both the classical gravitational field and its quantum fluctuations. 

As a final remark, it should be stressed  that the energy-extraction mechanism cannot take place when $k_1$ is negative. Indeed,  in this case $g_{tt}$ is always negative when $r>r_3$  and  the horizon located at $r=r_3$ is hidden by the null hypersurface $r=\tilde{r}_3$ (see Eq. \eqref{inequality-tilde-r3-r3} and Fig. \ref{Fig-k1-negative}).

\subsection{Curvature invariants} \label{Sec:curvature-inv}

Starting from the metric \eqref{Schwarzschild_metric_quantum-standard}, we find that the nonvanishing Christoffel symbols of the quantum Schwarzschild geometry read as
\begin{subequations}
\label{Christoffel-symbols}
\begin{align}
\Gamma^r_{rr} &=\frac{\RS  \left[9 k_1 \LP^2 (r-\RS)^2-2 r^4\right]}{2 r (r-\RS) \left[2 r^4+3 k_1 \LP^2  \RS \left(\RS-r\right)\right]}= \frac{-\RS}{2r \left(r-\RS\right)} + \left[\frac{9}{4}k_1 \frac{\RS \LP^2}{r^4} +\OR{-5}\right],
\\
\Gamma^r_{\theta \theta} &=\frac{2 r^4 (\RS-r)}{2 r^4+ 3 k_1 \LP^2 \RS \left(\RS-r\right) }= R_{\rm S} - r -\left[\frac{3}{2}k_1 \frac{ R_{\rm S} \ell_{\rm P}^2}{r^2}+\OR{-3}\right],
\\
\Gamma^r_{\phi \phi} &=\frac{2 r^4  (\RS-r) \sin^2\theta }{2 r^4+ 3 k_1 \LP^2 \RS \left(\RS-r\right)}= \left(\RS-r\right) \sin^2 \theta -\left[\frac{3}{2}k_1 \frac{ R_{\rm S} \ell_{\rm P}^2}{r^2} \sin^2 \theta + \OR{-3}\right],
\\
\Gamma^r_{tt} &=\frac{\RS (r-\RS) \left(2 r^2-3 k_1 \LP^2\right)}{2 r \left[2 r^4+ 3 k_1 \LP^2 \RS \left(\RS-r\right)\right]}= \frac{\RS \left(r-\RS \right)}{2 r^3} -\left[\frac{3}{4} k_1 \frac{ R_{\rm S} \ell_{\rm P}^2}{r^4}+\OR{-5}\right], 
\\
\Gamma^{\theta}_{\theta r} &=\Gamma^{\phi}_{\phi r} =\frac{1}{r},
\\
\Gamma^{\theta}_{\phi \phi} &=- \sin \theta \cos \theta,
\\
\Gamma^{\phi}_{\phi \theta} &= \cot \theta,
\\
\Gamma^t_{tr} &=\frac{\RS \left(2 r^2-3 k_1 \LP^2\right)}{4r^4+2r \RS \left(k_1 \LP^2-2r^2\right)}=\frac{\RS}{2r \left(r-\RS\right)} -\left[ \frac{3}{4}k_1 \frac{\RS \LP^2}{r^4} + \OR{-5}\right],
\end{align}
\end{subequations}
where in the last equality we  have retained only the leading-order quantum corrections inside the square brackets.
It is worth mentioning  that  $\Gamma^r_{rr} \neq -\Gamma^t_{tr}$, unlike the classical case. 

We can thus calculate the curvature invariants
\begin{subequations}
\label{curvature-inv}
 \begin{align}
 R^{\alpha \beta \gamma \delta} R_{\alpha \beta \gamma \delta} &= \frac{12 \RS^2}{r^6}  -60 k_1 \frac{\LP^2 \RS^2}{r^8} + \OR{-9},
 \\
 R^{\mu \nu} R_{\mu \nu} &= 9 k_1^2 \frac{\RS^2 \LP^4}{r^{10}}+\OR{-11},
 \\
 R&=g^{\mu \nu}R_{\mu \nu}=3 k_1 \frac{\RS \LP^2}{r^5}+\OR{-6},
 \end{align}   
\end{subequations}
where we have neglected   quantum corrections beyond  the leading order. These relations show that  $r=0$ represents a curvature singularity.  Moreover, curvature invariants are finite both at  the horizons defined in Sec. \ref{Sec:horizon} and the null hypersurfaces examined in Sec. \ref{Sec:analysis-g-rr}. Therefore, they  merely represent coordinate singularities.

\subsection{Geodesic equations}\label{Sec:analysis-geod}

It is well-known that freely falling particles follow a geodesic path \cite{Wald,Weinberg1972}. This can be worked out by means of Eq. \eqref{Christoffel-symbols}. Exploiting standard arguments  (see e.g. Ref. \cite{Weinberg1972}\footnote{See also Refs. \cite{Hackmann2008,Cieslik2022,Battista2022a,Fathi2022} for some recent studies of the geodesic motion framed in  Schwarzschild geometry and some extensions thereof.}), we easily find that the geodesic motion in the quantum geometry \eqref{Schwarzschild_metric_quantum-standard} is ruled by the following set of equations (for simplicity, we hereafter set $c=1$)
\begin{subequations}
\label{geod-equations-1}
\begin{align}
 & r^2 \frac{\dd \phi}{\dd t}   = J B(r),
 \\
& \frac{A(r)}{B^2(r)} \left(\frac{ \dd r}{\dd t}\right)^2 + \frac{J^2}{r^2}-\frac{1}{B(r)} =-E,
 \\
&  \dd \tau^2 = E B^2(r) \dd t^2,
\end{align}
\end{subequations}
while 
the orbit shape is described by
\begin{align}
\label{geod-equations-2}
\frac{A(r)}{r^4} \left(\frac{ \dd r}{\dd \phi}\right)^2 + \frac{1}{r^2} -\frac{1}{J^2 B(r)}=-\frac{E}{J^2},
\end{align}
where $\tau $ is the proper time, and $E$ and $J$ are constants of motion (in particular, $E$ is positive for massive particles and  vanishes for photons). 

As a first application of the geodesic dynamics, in Sec. \ref{Sec:effective-potential} we consider timelike geodesics and work out  the form of the effective potential, which permits to establish the radius of the innermost stable circular orbit (ISCO). Then,  the bending of light effect on null geodesics is studied in Sec. \ref{Sec:bending-light}.

\subsubsection{Effective potential and innermost stable circular orbit }\label{Sec:effective-potential}

Starting from Eq. \eqref{geod-equations-1}, the equation pertaining to timelike geodesics can be put in the form 
\begin{align}
 \frac{1}{2} \dot{r}^2 + \frac{1}{2} \frac{1}{A(r)} \left(\frac{L^2}{r^2}+1\right)   = \frac{1}{2} \frac{\mathscr{E}^2}{A(r)B(r)},
 \label{geod-timelike-1}
\end{align}
where a dot stands for differentiation with respect to the proper time, and 
\begin{align}
 L^2  &:= \frac{J^2}{E},   
 \nonumber \\
 \mathscr{E}^2 &:= \frac{1}{E},
\end{align}
are the constants of motion associated to the invariance under    spatial rotations and time translations, respectively,  of the metric \eqref{Schwarzschild_metric_quantum-standard}. Upon using Eq. \eqref{A-and-B-of-r-quantum} and neglecting $\OR{-4}$ terms, Eq. \eqref{geod-timelike-1} gives
\begin{align}
    \frac{1}{2} \dot{r}^2 + V_{\rm eff}(r)  = \frac{1}{2} \mathscr{E}^2, 
\end{align}
 the quantum corrected effective potential being
\begin{align}
V_{\rm eff}(r) = \frac{1}{2} \left(1-\frac{\RS}{r}\right) \left(\frac{L^2}{r^2}+1\right) + \frac{1}{2} k_1 \left(\frac{3}{2} - \mathscr{E}^2 \right) \frac{\RS \LP^2}{r^3},    
\label{effective-potential}
\end{align}
where we can see  that the new quantum term goes like $r^{-3}$ and depends, in particular,  on $\mathscr{E}^2$. 

The first order derivative of $V_{\rm eff}(r)$ is
\begin{align}
    V^\prime_{\rm eff}(r)= \frac{1}{r^4} \left[\frac{r^2 \RS}{2} + L^2 \left(\frac{3}{2} \RS-r \right) + \frac{3}{4} k_1 \RS \LP^2 \left(2 \mathscr{E}^2 -3\right)\right],
\end{align}
and hence its extrema read as
\begin{align}
\mathcal{R}_{\mp }   = \RS^{-1} \left[ L^2 \mp \sqrt{ L^4 - 3L^2 \RS^2 - \frac{9}{2} k_1 \RS^2 \LP^2  \left( \frac{2}{3} \mathscr{E}^2-1\right)}\right]. 
\end{align}
These turn out to be strictly positive real-valued quantities provided that the following conditions are satisfied: 
\begin{subequations}
\label{system-1}
\begin{align}[left = \empheqlbrace\,]
& L^2 \leq L^2_{-} \cup L^2 \geq L^2_{+}, \label{L-2-and-L-mp}\\
& L^2 > L^2_{\star},
\end{align}
\end{subequations}
where
\begin{subequations}
\begin{align}
L^2_{\mp} &:= \frac{3}{2} \RS^2 \left[ 1 \mp \sqrt{1+2 k_1 \frac{\LP^2}{\RS^2} \left(\frac{2}{3} \mathscr{E}^2-1\right)} \right],
\\
L^2_{\star}&:= - \frac{3}{2} k_1 \LP^2 \left(\frac{2}{3} \mathscr{E}^2-1\right).
\end{align}
\end{subequations}
Notice that   $\mathcal{R}_{-}=\mathcal{R}_{+}$ when $L^2 = L^2_{\mp}$, and  both $L^2_{-}$ and $L^2_{\star}$ vanish in the classical limit.

It is easy to check that the parameters  $L^2_{\mp}$ are real and strictly positive if
\begin{subequations}
\label{system-2}
\begin{align}[left = \empheqlbrace\,]
&1+2k_1 \frac{\LP^2}{\RS^2} \left(\frac{2}{3} \mathscr{E}^2-1\right) > 0,
\label{relation-3}\\
& k_1 \left(\frac{2}{3} \mathscr{E}^2-1\right) <0, \label{relation-4}
\end{align}
\end{subequations}
where in Eq. \eqref{relation-3} we have considered the symbol \qm{$>$} instead of  \qm{$\geq$} to guarantee that $L^2_{-}$ and $L^2_{+}$ differ; moreover, we note that $L^2_{\star}>0$ owing to Eq. \eqref{relation-4}.

The resolution of the systems \eqref{system-1} and \eqref{system-2} in the case of negative $k_1$ yields
\begin{subequations}
\begin{empheq}[left={k_1<0 \Rightarrow \empheqlbrace\,}]{align}
& \frac{3}{2} <  \mathscr{E}^2 < \frac{3}{2} \left[1+ \frac{\RS^2}{\LP^2 (-2k_1)}\right], \\
& L^2 \geq L^2_{+}, 
\label{L-squared-greater-L-plus}
\end{empheq}
\end{subequations}
whereas for positive $k_1$ we have
\begin{subequations}
\label{conditions-energy-ang-mom-positive-k1}
\begin{empheq}[left={k_1>0 \Rightarrow \empheqlbrace\,}]{align}
& \frac{3}{2} \left[1- \frac{\RS^2}{\LP^2 (2k_1)}\right] <  \mathscr{E}^2 < \frac{3}{2}, \label{energy-squared-positive-k1} \\
& L^2_{\star} < L^2 \leq L^2_{-} \cup L^2 \geq L^2_{+}, 
\label{L-squared-smaller-L-minus}
\end{empheq}
\end{subequations}
where we have taken into account that   $L^2_{-} < L^2_{\star} < L^2_{+}$ if $k_1 <0$, while
$L^2_{\star} <  L^2_{-}$ when $k_1 >0$. We also point out that Eq. \eqref{energy-squared-positive-k1} does not imply that $\mathscr{E}^2$ is positive. 

\begin{figure}[bht!]
\centering\includegraphics[scale=0.70]{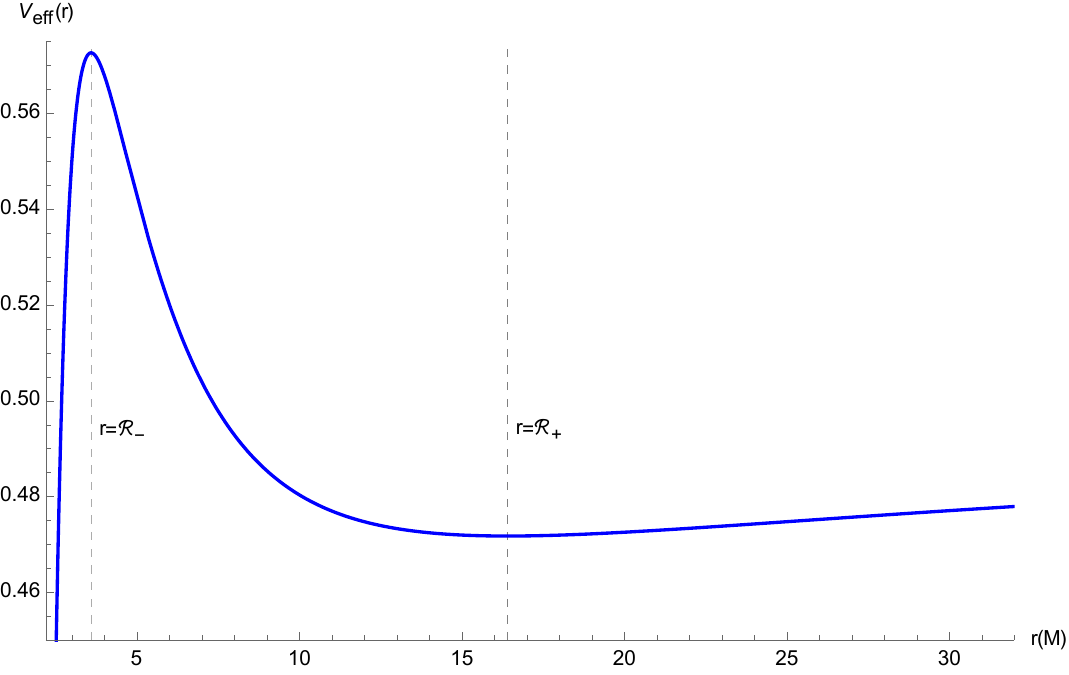}\hspace{1.25cm}
\caption{The effective potential \eqref{effective-potential}. The following values have been chosen: $G=c=\hbar=1$, $k_1=0.5$, $M=1$, $\mathscr{E}^2 =0.8$, and $L^2 = 20 M^2$,  the last two values satisfying Eq. \eqref{conditions-energy-ang-mom-positive-k1}. The  dashed grey lines are located at $r=\mathcal{R}_{-}$ and $r=\mathcal{R}_{+}$, with  $\mathcal{R}_{-}=3.6 M$ and $\mathcal{R}_{+}=16.4 M$. }
\label{Fig-potential}
\end{figure}

Thanks to the above conditions,    $\mathcal{R}_{-}$ is a maximum of the effective potential \eqref{effective-potential}, while $\mathcal{R}_{+}$  a minimum, like in the classical scenario  (see Fig. \ref{Fig-potential}). Therefore, stable circular orbits  exist at the radius $r=\mathcal{R}_{+}$.
The ISCO radius is defined as the minimum value of $\mathcal{R}_{+}$. This  is attained  when  $L^2 = L^2_{+}$ if $k_1<0$, and   $L^2 = L^2_{-}$ if $k_1>0$  (cf. Eqs. \eqref{L-squared-greater-L-plus} and \eqref{L-squared-smaller-L-minus}). In this way, we obtain
\begin{align}
\mathcal{R}_1^{\rm ISCO} &= \frac{L^2_{+}}{\RS} = \frac{3}{2} \RS \left[1+\sqrt{1+2 k_1 \frac{\LP^2}{\RS^2} \left(\frac{2}{3} \mathscr{E}^2 -1\right)} \right],  \qquad  (k_1 <0),
\label{ISCO-1}
\\
\mathcal{R}_2^{\rm ISCO} &=\frac{L^2_{-}}{\RS} = \frac{3}{2} \RS \left[1-\sqrt{1+2 k_1 \frac{\LP^2}{\RS^2} \left(\frac{2}{3} \mathscr{E}^2 -1\right)} \right],   \qquad (k_1 >0).
\label{ISCO-2}
\end{align}
As a consequence of Eq.  \eqref{system-2}, the modulus of the quantum term appearing in Eqs. \eqref{ISCO-1} and \eqref{ISCO-2} is less than $1$. Then, bearing in mind Eq. \eqref{Rs-bigger-than-lp}, we can expand the square root and hence   the above relations yield
\begin{align}
\mathcal{R}_1^{\rm ISCO} &= 3 \RS -\frac{L^2_{\star}}{\RS} + \OO\left(\LP^4/\RS^3\right),  \qquad  (k_1 <0),
\label{ISCO-1-expanded}
\\
\mathcal{R}_2^{\rm ISCO} &=\frac{L^2_{\star}}{\RS}+ \OO\left(\LP^4/\RS^3\right),   \qquad \quad \qquad (k_1 >0).
\label{ISCO-2-expanded}
\end{align}
It is worth stressing that Eq. \eqref{ISCO-2-expanded} is a consequence of the fact that $L^2_{-} = L^2_{\star} + \OO \left(\LP^4/\RS^2\right)$;  furthermore, it is easy to show that $\mathcal{R}_2^{\rm ISCO} < \tilde{r}_1$ if we take into  account Eq. \eqref{energy-squared-positive-k1} and allow for the reasonable hypothesis $6 k_1 \LP^2 / \RS^2 < 1$ (cf. Eq. \eqref{Rs-bigger-than-lp}). This   means that the ISCO having radius $r=\mathcal{R}_2^{\rm ISCO}$ cannot be observed from outside the null hypersurface located at $r=\tilde{r}_1$. 

From the  analysis of this section, we can deduce that the setup with $k_1 >0$ is plagued by an important issue, as Eq. \eqref{ISCO-2-expanded} greatly differs from the classical result. This seems to be against the spirit of the EFT framework, which relies on the assumption that quantum corrections should slightly affect  the classical contributions.  On the other hand, when $k_1 <0$  Eq. \eqref{ISCO-1-expanded} naturally respects the expectations  of the EFT scheme.

\subsubsection{Bending of light  } \label{Sec:bending-light}

The knowledge of the geodesic equations permits to analyze  the  deflection of light in the Schwarzschild geometry.  Starting from Eqs. \eqref{geod-equations-1} and \eqref{geod-equations-2}, it can be shown that the trajectory of a photon approaching the body of mass $M$ is given by
\begin{align}
 \phi(r)=\phi_\infty + \int  \limits^\infty_r \frac{\dd r^\prime}{r^\prime} \sqrt{A(r^\prime)\left[\frac{r^{\prime 2}}{r_0^2}\frac{B(r_0)}{B(r^\prime)}-1\right]^{-1}},
\end{align}
and the bending angle is
\begin{align}
\Delta \phi = 2\vert \phi(r_0)-\phi_\infty \vert - \pi,     
\end{align}
 $\phi_\infty$ and $r_0$  being the incident direction and the distance of closest approach, respectively (see Fig. 8.1 in Ref. \cite{Weinberg1972} for further details).  Bearing in mind Eq. \eqref{A-and-B-of-r-quantum},  we find, after expanding in the small parameters $\RS/r$ and $\RS/ r_0$,
 \begin{align}
\phi(r)-\phi_\infty &=     \int  \limits^\infty_r \frac{\dd r^\prime}{r^\prime \sqrt{r^{\prime 2}/r_0^2-1}} \Biggl[ 1+ \frac{1}{2} \frac{\RS}{r^{\prime}}+\frac{1}{2} \frac{\RS r{^\prime}}{r_0(r^\prime+r_0)}+\frac{3}{4} \frac{\RS^2 }{r_0(r^\prime+r_0)}+\frac{3}{8} \frac{\RS^2 r^{\prime 2} }{r_0^2(r^\prime+r_0)^2}
\\ \nn 
&+\frac{3}{8} \frac{\RS^2 } {r^{\prime 2}} + \frac{15}{16} \frac{\RS^3}{ r^{\prime }  r_0 (r^\prime+r_0)}+\frac{15}{16} \frac{\RS^3 r^{\prime }}{ r_0^2   (r^\prime+r_0)^2}+\frac{5}{16} \frac{\RS^3 r^{\prime 3}}{ r_0^3   (r^\prime+r_0)^3} + \frac{5}{16} \frac{\RS^3}{r^{\prime 3}}
\\ \nn
&-\frac{k_1}{4} \frac{\RS \LP^2 \left(r^{\prime 2}+r^\prime r_0 + r_0^2\right)}{r^\prime r_0^3 (r^\prime + r_0)}- \frac{3 k_1}{4} \frac{\RS \LP^2}{r^{\prime3}}+\dots \Biggr],
 \end{align}
which yields a deflection angle
\begin{align}
\Delta \phi = 2 \frac{\RS}{r_0} + \frac{\RS^2}{r_0^2} \left(\frac{15 \pi}{16} -1\right)    +\frac{\RS^3}{r_0^3} \left(\frac{61}{12}-\frac{15 \pi}{16} \right) -2  k_1  \frac{\RS \LP^2}{r_0^3} + {\rm O}\left(r_0^{-4}\right),  
\label{deflection-angle-quantum}
\end{align}
the second-order and third-order classical post-Newtonian corrections being in agreement with the results given in   Refs. \cite{Epstein1980,Fischbach1980,Richter1982,Bodenner2003,Rodriguez2017}. For a light ray grazing the surface of the Sun we have $r_0 = 6.96 \times 10^{8}$ m, and hence
\begin{align}
\Delta \phi_{\odot}= 1.75^{\prime \prime}+ \left(7.2 \times 10^{-6}\right)^{\prime \prime }+\left(3.4 \times 10^{-11}\right)^{\prime \prime } -  \left(4.6 \, k_1 \times 10^{-93}\right)^{\prime \prime },
\end{align}
which clearly shows that the quantum correction is undetectable. 

The quantum contributions occurring in Eq. \eqref{deflection-angle-quantum} have the same formal structure as those of Refs. \cite{Bohr2015,Bohr2016,Bai2017,Chi2019}, although the numerical coefficients are different.  This can be explained with the fact that  we have exploited standard general-relativity tools and we have assumed  that light consists of photons  travelling along a classical geodesic trajectory. On the other hand,  the framework adopted in Refs. \cite{Bohr2015,Bohr2016,Bai2017,Chi2019} makes use of quantum techniques to compute one-loop scattering amplitudes which allow for the  violation of the equivalence principle. However, as pointed  out before, the final outcome of Refs. \cite{Bohr2015,Bohr2016} does not agree with the one reported in Refs. \cite{Bai2017,Chi2019}. Indeed,  some discordant results can be found  
in the literature devoted to EFT models of gravity  due to the underlying laborious calculations. This means that feasible experiments regarding observable phenomena should  be conceived  to ascertain the most correct model  pertaining to  quantum-gravity effects at low energies.

\section{Concluding remarks}\label{Sec:Conclusions}

GR is  not perturbatively renormalizable. Despite that,  its low-energy domain can be isolated by integrating out the high-energy degrees of freedom. This comes as a consequence of the fact that GR can be naturally treated as an  EFT where the Planck energy serves as a cutoff scale. Therefore, explicit field-theory calculations can be carried out and low-energy quantum predictions  can be made without the knowledge of the  high-energy regime of the (unknown) theory of quantum gravity. 

In this paper, we have studied the quantum corrected version of the Schwarzschild geometry by exploiting the EFT paradigm. The metric has been written in Schwarzschild coordinates $x^\mu=(ct,r,\theta,\phi)$ (see Eqs. \eqref{Schwarzschild_metric_quantum-standard} and \eqref{A-and-B-of-r-quantum}) by constructing  the explicit coordinate transformation relating the harmonic coordinates $X^\mu$ with $x^\mu$ (see Sec. \ref{Sec:Quantum-corrected-Schwarzschild-metric}). Then, in Sec. \ref{Sec:horizon} we have worked out the metric horizon(s), while  the analysis of the  hypersurfaces having constant radius is contained in Sec. \ref{Sec:analysis-g-rr}.  A possible energy-extraction process which violates the NEC  has been discussed in Sec. \ref{Sec:energy-extraction} and   the curvature invariants have been dealt with in Sec. \ref{Sec:curvature-inv}. Subsequently, we have evaluated the timelike geodesics along with the related quantum effective potential and ISCO orbits (see  Sec. \ref{Sec:effective-potential}). Lastly, we have considered the bending of light in Sec. \ref{Sec:bending-light}. 

In our investigation, we have found that the quantum metric does not satisfy the condition $-g_{tt} = g^{rr}$, unlike some models proposed in the literature (see e.g. Refs. \cite{Bonanno2000,Bargueno2016}). Moreover, the low-energy quantum corrections are kept in their most general form, since they have been parametrized by the constant $k_1$. The solution having positive $k_1$ is characterized by the presence of two horizons and two null hypersurfaces if $M > \sqrt{3}  M^\star $ (cf. Eqs. \eqref{solution-r-1}, \eqref{solution-r-2}, \eqref{solution-r-tilde-1}, and \eqref{solution-r-tilde-2}); these disappear  as soon as $M<  M^\star$, which thus gives rise to a regime where  $r=0$ represents a naked singularity. In this limit, the EFT approach comes to naught, since the  quantum contributions   would become as important as the classical ones.  Despite that, these features might be linked to a  conjecture of   Hawking, who has suggested that black holes with masses lower than the Planck mass and  radii smaller than the Planck length could not be created \cite{Hawking1971}. However, the setup having positive $k_1$ seems to  suffer from some shortcomings. First of all, the radii of the  horizon located at $r=r_2$  and the hypersurface $r=\tilde{r}_2$ can be smaller than the Planck length (see Eqs. \eqref{second-horizon} and \eqref{solution-r-tilde-2}). Nevertheless, we have seen that this drawback can be sorted out thanks to the presence  of the   null hypersurface $r=\tilde{r}_1$. Furthermore, the solution with $k_1 >0$ leads to the energy-extraction mechanism set out in Sec. \ref{Sec:energy-extraction}, which has no classical analogue. In addition,  the   ISCO radius \eqref{ISCO-2-expanded} deviates significantly from the classical expectations. On the other hand,  the scenario with $k_1<0$ predicts that: $(i)$ for any real-valued $M$, one  horizon and one null hypersurface only  are present (see Eqs.  \eqref{quantum-horizon} and \eqref{solution-r-tilde-3}); $(ii)$  the ISCO radius \eqref{ISCO-1-expanded} is close to the classical one. Despite  the rationality of $(ii)$, there exists a regime where the result $(i)$ can be contrary to the spirit of EFTs. In fact, since there is no lower bound on $M$ when $k_1<0$ (cf. Eq. \eqref{case-k1-negative}),  Eqs. \eqref{Rs-bigger-than-lp} and \eqref{M-bigger-M-Planck} are not automatically satisfied and hence the occurrence of primordial black holes cannot be avoided. On the other hand, this does not happen in the framework with $k_1$  positive, where the horizons \eqref{solution-r-1} and \eqref{solution-r-2} and the null hypersurfaces \eqref{solution-r-tilde-1} and \eqref{solution-r-tilde-2} can be defined only if $M \gtrsim M_{\rm P}$.

On the basis of our analysis, it seems that there is no preferred choice  of sign of $k_1$ for which the main features of the classical Schwarzschild geometry are slightly affected by the low-energy quantum corrections, although  at first sight and modulo the problems due to the lack of bounds on $M$, one might conclude  that the case with negative $k_1$  is  the most suitable one for describing the quantum Schwarzschild metric.  The absence of a clear and reasonable alternative between the options  $k_1 <0$ and $k_1 >0$ ties in with the outcome of Ref. \cite{Battista2014a},  where it has been proved that when EFT techniques are applied to the examination of the restricted three-body problem, some changes of its qualitative properties with respect to Newtonian theory are  always unavoidable. Furthermore, it is worth mentioning that  it has been demonstrated  \cite{Battista2015b,Battista-book2017}  how the restricted three-body problem permits to obtain a  criterion to determine the most appropriate coefficients $k_1,k_2,k_3$ appearing in the quantum corrected potential (cf. Eqs. \eqref{constants-k} and \eqref{quantum-coorected-potential}). 

Another remark is in order. As we have explained before,  we have obtained a quantum correction to the light bending which differs from the one displayed in  Refs.  \cite{Bohr2015,Bohr2016,Bai2017,Chi2019} (see Eq. \eqref{deflection-angle-quantum}). Although this discrepancy is  due to the distinct pattern employed in this manuscript, it should be stressed that even Refs. \cite{Bohr2015,Bohr2016} and \cite{Bai2017,Chi2019} disagree. This is not a novel facet of the literature devoted to EFTs, as also many calculations regarding  the quantum Schwarzschild metric and Newtonian potential are not tantamount (see e.g. Table 1 in Ref. \cite{Bargueno2016}). These circumstances   motivate us to further explore the potentialities and the limits of EFTs. 

This paper can open up some interesting future perspectives. First of all, a more  detailed study of the geodesic motion along the lines of Refs. \cite{Chandrasekhar1985,Hackmann2008,Cieslik2022,Battista2022a,Fathi2022} should be pursued. Moreover, it would be interesting to consider other metrics, such as the Kerr, Reissner-Nordstr\"{o}m, and Kerr-Newman ones, by exploiting the results of Refs. \cite{Donoghue2001,Bohr2003}. These topics deserve to be addressed in a separate paper.

\section*{Acknowledgements}
This work  is supported by the Austrian Science Fund (FWF) grant P32086. It is a pleasure to thank G. Esposito for having carefully read the preliminary version of this work. The paper is dedicated to my loved one Federica.

\bibliography{references}
\end{document}